\documentclass[conference]{IEEEtran}
\IEEEoverridecommandlockouts
\usepackage[english]{babel}
\usepackage{amsthm}
\usepackage{titlesec}
\usepackage[ruled,vlined]{algorithm2e}
\usepackage{algcompatible}
\usepackage{graphicx}
 
\usepackage{float}
\usepackage{ragged2e}
\usepackage[labelfont=bf,font=small,justification=justified,singlelinecheck=false]{caption}
\usepackage{etoolbox}
\usepackage{subcaption}
\makeatletter
\patchcmd{\@makecaption}{\centering}{\justifying}{}{}
\makeatother
\usepackage{tikz}
\usepackage{array}
\usepackage{changepage}
\usepackage[utf8]{inputenc}
\usepackage{pgfplots} 
\usepackage{pgfgantt}
\usepackage{pdflscape}
 \usepackage{relsize}
\usepackage[export]{adjustbox}
\pgfplotsset{compat=newest} 
\pgfplotsset{plot coordinates/math parser=false}
\pgfplotsset{compat=1.18}
\usepackage{pgfplots}
\usetikzlibrary{spy}

\usepackage{cite}
\usepackage{amsmath,amssymb,amsfonts}
\usepackage{dsfont} 
\usepackage{stfloats}
\newtheorem{innercustomgeneric}{\customgenericname}
\providecommand{\customgenericname}{}
\newcommand{\newcustomtheorem}[2]{%
  \newenvironment{#1}[1]
  {%
   \renewcommand\customgenericname{#2}%
   \renewcommand\theinnercustomgeneric{##1}%
   \innercustomgeneric
  }
  {\endinnercustomgeneric}
}

\newcustomtheorem{customthm}{Theorem}
\newcustomtheorem{customlemma}{Lemma}
\newcustomtheorem{customprop}{Proposition}
\newcustomtheorem{customcor}{Corollary}
\newcustomtheorem{customrem}{Remark}
\usepackage{lipsum}
\usepackage{amsmath}
\usepackage[nolist,printonlyused]{acronym} 
\usepackage{amssymb}
\usepackage{mathtools}
\usepackage{url}
\usepackage{graphicx}  
\usepackage{float}  

\newcommand{\yy}{\mathbf{y}}
\newcommand{\xx}{\mathbf{x}}
\newcommand{\uu}{\mathbf{u}}

\newcommand{\hh}{\mathbf{h}}
\newcommand{\nn}{\mathbf{n}}
\newcommand{\vv}{\mathbf{v}}
\newcommand{\pp}{\mathbf{p}}

\newcommand{\Mbs}{{M_{\rm{BS}}}}

\newcommand{\Mue}{M_{\rm{UE}}}

\newcommand{\bs}{{\normalfont{\textrm{BS}}}}
\newcommand{\ue}{{\normalfont{\textrm{UE}}}}

\newcommand{\rmd}{{\normalfont{\textrm{D}}}}

\newcommand{\OOmega}{\boldsymbol{\Omega}}

\newcommand{\rmp}{{\textrm{p}}}

\newcommand{\UU}{\mathbf{U}}

\newcommand{\YY}{\mathbf{Y}}
\newcommand{\NN}{\mathbf{N}}
\newcommand{\HH}{\mathbf{H}}

\newcommand{\WW}{\mathbf{W}}
\newcommand{\QQ}{\mathbf{Q}}
\newcommand{\FF}{\mathbf{F}}

\newcommand{\DD}{\mathbf{D}}
\newcommand{\CC}{\mathbf{C}}

\newcommand{\diag}{\text{diag}}

\newcommand{\hermit}{\mathsf{H}}

\usepackage{accents}

\newcommand{\TT}{\mathbf{T}}

\newcommand{\MM}{\mathbf{M}}

\newcommand{\ptot}{{p_\textrm{tot}}}
\newcommand{\AAb}{\mathbf{A}}

\newcommand{\norm}[1]{\left\lVert#1\right\rVert}

\newcommand{\ee}{\mathbf{e}}

\newcommand{\vecc}{\text{vec}}

\newcommand{\EE}{\mathbf{E}}

\newcommand{\dl}{ {\textrm{DL}}}
\newcommand{\ul}{ {\textrm{UL}}}

\acrodef{SISO}[SISO]{single-input single-output}
\acrodef{PRB}[PRB]{physical resource block}
\acrodef{AP}[AP]{access point}
\acrodef{AR}{autoregressive}
\acrodef{UE}[UE]{user equipment}
\acrodef{ULA}[ULA]{uniform linear array}
\acrodef{ML}[ML]{maximum likelihood}
\acrodef{CPU}[CPU]{central processing unit}
\acrodef{DL}{downlink}
\acrodef{UL}{uplink}
\acrodef{FPP}[FPP]{Feasible-point pursuit}
\acrodef{LoS}[LoS]{line of sight}
\acrodef{NLoS}[NLoS]{non-line-of-sight}
\acrodef{RCS}[RCS]{radar cross section}
\acrodef{AoD}[AoD]{angle of departure}
\acrodef{ZF}[ZF]{zero-forcing}
\acrodef{AoA}[AoA]{angle of arrival}
\acrodef{CRB}[CRB]{Cram\'er-Rao bound}
\acrodef{FIM}[FIM]{Fisher information matrix}
\acrodef{AN}[AN]{artificial noise}
\acrodef{SINR}[SINR]{signal-to-interference-plus-noise ratio}
\acrodef{TDD}[TDD]{time-division duplexing}
\acrodef{SNR}[SNR]{signal-to-noise ratio}
\acrodef{QoS}[QoS]{quality of service}
\acrodef{SDR}[SDR]{semi-definite relaxation}
\acrodef{SDP}[SDP]{semi-definite program}
\acrodef{ISAC}[ISAC]{integrated sensing and communications}
\acrodef{PLS}[PLS]{physical layer security}
\acrodef{SIC}[SIC]{successive interference cancellation}
\acrodef{CSI}[CSI]{channel state information}
\acrodef{CSIT}[CSIT]{channel state information at the transmitter}
\acrodef{CSIR}[CSIR]{channel state information at the receiver}
\acrodef{MUI}[MUI]{multi-user interference}
\acrodef{RIS}[RIS]{Reconfigurable intelligent surface}
\acrodef{AO}[AO]{alternating optimization}
\acrodef{SIMO}[SIMO]{Single Input Multiple Output}
\acrodef{MISO}[MISO]{multiple-intput single output}
\acrodef{MIMO}[MIMO]{multiple-input multiple-output}
\acrodef{MU}{multi-user}
\acrodef{BS}[BS]{base station}
\acrodef{CEE}[CEE]{channel estimation error}
\acrodef{CCP}[CCP]{convex-concave procedure}
\acrodef{MRT}[MRT]{ maximum-ratio transmission}
\acrodef{MRC}[MRC]{ maximum-ratio combining}
\acrodef{MM}[MM]{Minorization-Maximization}
\acrodef{PSD}[PSD]{positive semi-definite}
\acrodef{RZF}[RZF]{Regularized zero forcing}
\acrodef{CRZF}[CRZF]{Centralized regularized zero forcing}
\acrodef{RZF}[RZF]{regularized zero forcing}
\acrodef{LPZF}[LPZF]{Local Partial zero forcing}
\acrodef{LZF}[LZF]{Local zero forcing}
\acrodef{NF}[NF]{Near Field}
\acrodef{FF}[FF]{Far Field}
\acrodef{BD}[BD]{Beyond-diagonal}
\acrodef{OFDM}[OFDM]{orthogonal frequency division multiplexing}
\acrodef{MAPRT}[MAPRT]{maximum a-posteriori ratio test}
\acrodef{LRT}[LRT]{ likelihood ratio test}
\acrodef{CDF}[CDF]{Cumulative distribution function}
\acrodef{UPA}[UPA]{uniform planar array}
\acrodef{LMMSE}[LMMSE]{linear minimum mean square error}
\acrodef{MMSE}[MMSE]{minimum mean square error}
\acrodef{SE}[SE]{spectral efficiency}
\acrodef{CNR}[CNR]{clutter to noise ratio}
\acrodef{SCNR}[SCNR]{signal to clutter and noise ratio}
\acrodef{SOCP}[SOCP]{second order cone program}
\acrodef{TTD}[TTD]{true time delay}
\acrodef{PS}[PS]{phase shifter}
\acrodef{PEB}[PEB]{Position error bound}
\acrodef{JRC}[JRC]{Joint radar and communication}
\acrodef{GP}[GP]{gradient projection}
\acrodef{MO}[MO]{Manifold Optimization}
\acrodef{DEB}[DEB]{direction error bound}
\acrodef{REB}[REB]{range error bound}
\acrodef{LS}[LS]{least-squares}
\acrodef{MUSIC}[MUSIC]{MUltiple SIgnal Classification}
\acrodef{SR}[SR]{sum rate}
\acrodef{WMMSE}[WMMSE]{weighted-minimum mean square error}
\acrodef{BCD}[BCD]{block-coordinate descent}
\acrodef{SOCP}[SOCP]{second-order cone program}
\acrodef{DFT}[DFT]{discrete Fourier transform}
\acrodef{PSO}[PSO]{particle swarm optimization}
\acrodef{SIRV}[SIRV]{spherically-invariant random vector}
\acrodef{MF}[MF]{matched filter}
\acrodef{SCR}[SCR]{signal to clutter ratio}
\acrodef{CFAR}[CFAR]{constant false alarm rate}
\acrodef{RA}[RA]{range-angle}
\acrodef{RV}[RV]{range-velocity}
\acrodef{MU-MIMO}{multi-user multiple input multiple output}
\acrodef{NMSE}{normalized mean squared error}
\acrodef{SV}[SV]{singular value}
\acrodef{RMS}[RMS]{root mean square}

\begin{document}

\title{Geometry-Aware Channel Aging in MIMO Systems: When Is Kronecker Separability Enough?}

\author{
Steven~Rivetti,~\IEEEmembership{Student Member,~IEEE},
Gabor Fodor,~\IEEEmembership{Fellow,~IEEE}\\
Emil~Björnson,~\IEEEmembership{Fellow,~IEEE}, 
Mikael~Skoglund,~\IEEEmembership{Fellow,~IEEE}
  \thanks{
This work was supported by the SUCCESS project (FUS21-0026), funded by the Swedish Foundation for Strategic Research. G. Fodor was supported by the Swedish Strategic Research (SSF) FUS21-0004 SAICOM project.}
\thanks{The authors are with the School of Electrical Engineering and Computer Science (EECS), KTH Royal Institute of Technology, 11428 Stockholm, Sweden. Gabor Fodor is also with Ericsson Research, Stockholm, Sweden. } 
}

\maketitle

\begin{abstract}
This letter investigates when Kronecker-separable spatial correlation approximates the covariance of multiple-input-multiple-output (MIMO) channels. We propose a geometry-aware channel model in which the same propagation environment determines both the Weischelberger spatial coupling matrix and the temporal correlation coefficient. The propagation environment follows a 3GPP-inspired clustered channel model. We consider a MIMO system affected by symbol-level channel aging caused by the user equipment's mobility. We schedule both uplink and downlink pilots, thereby deriving the respective spectral efficiencies (SEs) in the presence of imperfect channel state information.
Numerical results show that channel aging's magnitude is strongly geometry-dependent, with diffuse propagation inducing greater temporal decorrelation and SE degradation than concentrated scattering. 
The results also identify regimes where Kronecker separability can be applied and regimes where geometry-aware Weischelberger coupling is needed.
\end{abstract}

\section{introduction}
\Ac{MIMO} systems rely on accurate \ac{CSI} to perform \ac{MU}-precoding and interference suppression. In practice, \ac{CSI} is acquired through pilot transmission and becomes progressively outdated over time due to \acp{UE} mobility and the dynamic evolution of the propagation environment.
This phenomenon, commonly referred to as channel aging\cite{fodor2023optimizing}, becomes prevalent in high-mobility scenarios, where the channel variations within the frame may become non-negligible\cite{Truong:13}, resulting in a degradation of the achievable \ac{SE}. 
The channel's temporal correlation decay is often modeled using Jake-like isotropic scattering models, which produce correlation coefficients mainly determined by the \ac{UE} velocity magnitude \cite{wang2003efficient}. Though mathematically tractable, this model fails to capture the effect of the propagation geometry, as \acp{UE} with the same velocity magnitude may exhibit different correlation decays depending on the positions of their dominant multipath clusters.

The propagation geometry is represented by the channel spatial covariance matrix. Thanks to its mathematical tractability, the Kronecker-separable model is among the most widely adopted \cite{bengtsson2006some}.  This model represents the MIMO channel covariance as the product of one-sided \ac{BS} and \ac{UE} correlation matrices. This separability assumption neglects the joint correlation between BS and UE, which is especially important in \ac{MU}-\ac{MIMO} networks. The Weischelberger \cite{weichselberger2006stochastic} model overcomes this limitation by introducing a coupling matrix that describes how power is distributed across BS-UE eigenmode pairs \cite{debbah2022uplink}. Nevertheless, this coupling is often treated as an abstract entity rather than being explicitly linked to the propagation geometry \cite{dovelos2020massive}. 

The need for a unified framework where the same propagation environment gives rise to both the spatial and the temporal correlation of the \ac{MIMO} channel motivates the main question in this letter: 
When is the Kronecker-separable model sufficiently accurate, and when is joint-correlation coupling modeling necessary to capture the propagation geometry?

\subsection{Contributions}
This letter proposes a geometry-aware MIMO 
channel-aging model that links spatial eigenmode coupling and temporal correlation through the same propagation geometry. 
The propagation geometry is inspired by the 3GPP-standardized cluster delay line channel model. The propagation environment is indeed made of propagation clusters, whose positions are randomly generated, even though it is worth mentioning that the same framework could be driven by ray tracing in practical deployments
We construct a Weischelberger coupling matrix by projecting the cluster-level covariance contributions onto the BS and UE eigenmodes. A cluster-dependent temporal correlation coefficient is then derived in which each cluster Doppler shift is weighted by its power, UE-side angle, and Doppler spread. To this end, we propose a Doppler-spread model that directly links the uncertainty in the angular domain to that in the Doppler domain.
Finally, we evaluate the symbol-level aging impact in a \ac{TDD} MIMO system with explicit \ac{UL} and \ac{DL} pilots, covariance-based channel estimation, and imperfect-\ac{CSI} \ac{SE} expressions. Numerical results identify regimes where the Kronecker approximation is accurate and regimes where geometry-aware coupling is needed, showing that diffuse propagation induces stronger temporal decorrelation and larger SE losses than concentrated scattering.

\section{System Model}
We consider a \ac{MU}-\ac{MIMO} system, where a \ac{BS}, located at the origin and equipped with $\Mbs$ antennas, serves $K$ \acp{UE}, equipped with $\Mue$ antennas, located at $\ee_k=[e_{k,x},e_{k,y}]^\top$.
The \ac{BS} performs \ac{OFDM} transmission around the carrier frequency $\overline{f}$, over $V$ active subcarriers with a subcarrier spacing of $\Delta_f$.
The subcarriers are partitioned into $B
$ blocks of $V_\textrm{cho}$ subcarriers, indexed by the block index $n$: the channel is assumed constant across subcarriers within the same block.
Here $T=1/\Delta_f + T_\textrm{cp} $ is the symbol duration, with $T_\textrm{cp}$ being the cyclic prefix duration.
We define the symbol time index $i$, with time unit $T$.
The frequency-domain \ac{UL} communication channel between \ac{UE} $k$ and \ac{BS} is denoted by $\HH_{k,i,n} \in \mathds{C}^{\Mbs \times \Mue}$.
\Ac{TDD} and \ac{UL}/ \ac{DL} channel reciprocity are assumed, i.e., $\HH_{k,i,n}^\dl =\HH_{k,i,n}^\top$.
We account for non-negligible time correlation between symbol times: the correlation decay is modelled as a first-order \ac{AR} process with a scaled identity state-transition matrix \cite{fodor2023optimizing}.
The channel's time correlation is thus defined as
   $ \mathds{E}[\hh_{k,i,n}\hh^\hermit_{k,i',n}]=\CC_k\zeta_k(|i-i'|),$
 where $\zeta_k(|i-i'|)$ captures the correlation decay. 
\begin{figure}[t!]
\begin{center}
   \resizebox{0.4\textwidth}{!}{
       \includegraphics[]{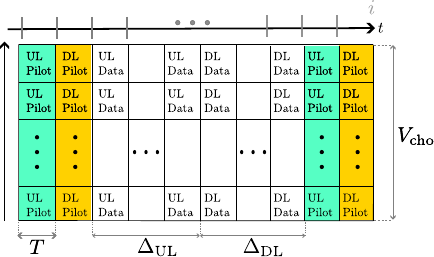}}
	  \caption{Representation of the pilot and data symbol transmission structure in the time-frequency domain. The communication channel is estimated at the BS and at each \ac{UE} thanks to the respective UL and DL pilots.}
       \label{pilot table}
		\end{center}
  \vspace{-8mm}
\end{figure}
The frequency domain \ac{UL} observation at the BS at time $i$ on a generic subcarrier $v$ belonging to block $n$ is defined as 
\vspace*{-2mm}
\begin{align}
\yy_{i,n}^\ul\triangleq\sum_{k=1}^K\alpha_k\HH_{k,i,n}\WW_{k,i,n} \boldsymbol{x}_{k,i,v}  + \nn_{i,v},
\end{align}
where  $\WW_{k,i,n} \in \mathbb{C}^{\Mue \times \Mue}$ is the precoding matrix adopted by \ac{UE} $k$ at time $i$ for all the subcarriers in block $n$.
Here, $\alpha_k$ accounts for the path loss and large-scale fading of UE $k$. 
Here $\xx_{k,i,v}\in \mathds{C}^{\Mue}$ are the transmitted symbols adopted by UE $k$ in its UL transmission.
We assume that  $\boldsymbol{x}_{k,i,v}\sim \mathcal{CN}(\mathbf{0},\mathbf{I}_{\Mue} )$.
The \ac{BS} receiver noise is $\nn_{i,v}\sim \mathcal{CN}(\mathbf{0},\mathbf{I}_{\Mue}\sigma_{\bs}^2)$.
Conversely, the frequency-domain \ac{DL} observation at \ac{UE} $k$  is defined as 
\begin{align}
\label{y generic data}
\yy_{k,i,n}^\dl
&\triangleq
\alpha_k\HH_{k,b.n}^\dl\FF_{i,n}\xx_{i,v} + \nn_{i,v},
\end{align}
\v
where $\FF_{i,n}=[\FF_{1,i,n},\dots,\FF_{K,i,n}] \in \mathds{C}^{\Mbs \times K\Mue}$ is the precoder matrix adopted for every subcarrier in the $n$-th block.
 The transmitted symbols  
are defined as $\xx_{i,v}=[\xx_{i,v,1},\dots,\xx_{i,v,K}]^\top\in\mathbb{C}^{\Mue K}$, with $\xx_{i,v,k}\sim \mathcal{CN}(\mathbf{0},\mathbf{I}_{\Mue})$.
Lastly, $\nn_{i,v,k}\sim \mathcal{CN}(\mathbf{0},\mathbf{I}_{\Mue}\sigma_{k}^2)$ is UE $k$ receiver noise.
\subsection{Weischelberger correlation model}
We here take the Weichselberger Rayleigh fading model \cite{weichselberger2006stochastic} as a reference and use it to study when the Kronecker-separable approximation is accurate. This model introduces an eigenmode coupling matrix, here denoted by $\OOmega_k \in \mathbb{R}_+^{\Mbs \times\Mue}$, specifying how the channel power is distributed across BS–UE eigenmode pairs, while the Kronecker model is obtained as a separable special case, where the BS and UE eigenmodes are perfectly uncoupled.
Let us define the one-sided correlation matrices of the UL channel between UE $k$ and BS as $\CC_{\bs,k}=\mathbb{E}[\HH_k\HH_k^\hermit] \in \mathbb{C}^{\Mbs \times\Mbs}$ and $\CC_{\ue,k}=\mathbb{E}[\HH_k^\dl\HH_k^{\dl,\hermit}] \in \mathbb{C}^{\Mue \times\Mue}$, with their respective eigendecompositions  $\CC_{\bs,k}=\UU_{\bs,k}\diag(\boldsymbol{\lambda}_{\bs,k})\UU_{\bs,k}^\hermit$ and $\CC_{\ue,k}=\UU_{\ue,k}\diag(\boldsymbol{\lambda}_{\ue,k})\UU_{\ue,k}^\hermit$.
The UL channel covariance $\CC_k^\ul=\mathds{E}[\vecc(\HH)\vecc(\HH)^\hermit]$ is  defined as 
\begin{align}\label{eq C_q ul}
    \CC_k^\ul\hspace{-1mm}=\hspace{-1mm}\left(\UU_{\ue,k}^* \otimes \UU_{\bs,k}\right)\DD_{\OOmega,k}\left(\UU_{\ue,k}^* \otimes \UU_{\bs,k}\right)^\hermit
\end{align}
Where $\DD_{\OOmega,k}=\diag(\vecc(\OOmega_k)).$
The DL channel covariance $\CC_k^\dl=\mathds{E}[\vecc(\HH^\dl)\vecc(\HH^\dl)^\hermit]$ follows by reciprocity: 
\begin{align}\label{eq C_q dl}
\CC_k^\dl\hspace{-1mm}=\hspace{-1mm}\left(\UU_{\bs,k}^* \otimes \UU_{\ue,k}\right)\DD_{\OOmega^\top\hspace{-1mm},k}\left(\UU_{\bs,k}^* \otimes \UU_{\ue,k}\right)^\hermit
\end{align}
For simplicity, the traces of both matrices are normalized to one. It should be noted that the coupling matrix $\OOmega_k^\textrm{kron}=\frac{1}{\Mbs\Mue} \boldsymbol{\lambda}_{\bs,k}\boldsymbol{\lambda}_{\ue,k}^\top$ yields a Kronecker-separable covariance model.
The covariance mismatch between the Kronecker-separable model and the jointly-correlated one is measured as 
\begin{align}
d_{\OOmega} = \max_{k} \frac{\norm{\OOmega_k - \OOmega^\textrm{kron}_k}^2_\textrm{F}}{\norm{\OOmega_k}^2_\textrm{F}}.    
\end{align}

\subsection{Propagation geometry}
The propagation environment between the BS and the UEs is composed of a set of $C$ scattering clusters. Cluster $c$ is centered at $\pp_c=[p_{x,c},p_{y,c}]^\top$ and is seen by the BS at the angle $\psi_c$ and at the angle $\varphi_{k,c}$ by UE $k$.
The angles of the rays scattered by each cluster follow a normal distribution with side-dependent angle spreads, denoted by $\varsigma_{\bs,c}$ and $\varsigma_{\ue,k,c}$.
Assume that the size of each cluster is described by its effective radius $r_c$. The angle spreads of those clusters are then computed as 
\begin{align}
\varsigma_{\bs,c}\hspace{-1mm}=\hspace{-1mm}\normalfont{\textrm{atan}}\left(\frac{r_c}{\norm{\pp_c}}\right), \varsigma_{\ue,k,c}=\normalfont{\textrm{atan}}\left(\frac{r_c}{\norm{\pp_c-\ee_k}}\right).
\end{align}
The rationale behind this model is that closer clusters scatter rays within a wider angular interval than farther ones.
The marginal covariances $\CC_{\bs,k}$ and $\CC_{\ue,k}$ are computed according to the model in \cite{demir2022channel}:  under the assumption of small angle spreads, we can write $\CC_{\bs}=\sum_{c=1}^C\rho_{c}^\textrm{BS} \boldsymbol{C}_{\bs,c}$, where 
\begin{align}
&\hspace{-2mm}\left[\boldsymbol{C}_{\bs,c}\right]_{l,m}\hspace{-1mm}\approx\hspace{0.5mm}e^{-j\pi (l-m)\sin(\psi_{c})}e^{-\frac{\pi^2}{2}((l-m) \cos(\psi_{c})\varsigma_{\bs,c})^2}.
\end{align}
The definition of $\CC_{\ue,k}$ follows from the same model and is omitted for brevity. 
Here $\rho_c^\textrm{BS}$ represents the power of cluster $c$ as seen from the BS; conversely, $\rho_{c,k}^\textrm{UE}$ represents the cluster power seen by UE $k$, while $\rho_{c,k}^\textrm{J}$ is the 
power of said cluster in the coupling matrix.
Each of those powers is generated by a geometry-based adaptation of the standardized power delay profile\cite{series2017guidelines}.
The clusters BS side, UE side, and joint delays are defined as 
$\overline{\tau}^\textrm{BS}_c=\norm{\pp_c}/c_0$, $\overline{\tau}^\textrm{UE}_{c,k}=\norm{\pp_c-\ee_k}/c_0$, and
$\overline{\tau}_{c,k}^\textrm{J}=\left(\norm{\pp_c}+\norm{\pp_c
-\ee_k} - \norm{\ee_k}\right)/c_0$, with $c_0$ denoting the speed of light. 
Each of those delays is then re-normalized such that the closest cluster has zero delay, i.e., $\tau_{c,k}=\overline{\tau}_{c,k}- \min_l \overline{\tau}_{l,k}$.
The cluster powers are then computed as \cite[p.~54]{series2017guidelines}
\begin{align}
    \rho_{c,k}=\exp\left\{ - \tau_{c,k} (r_\tau-1)/(r_\tau DS)\right\}10^{-Z_{c,k}/10},
\end{align}
where $DS$ and $r_\tau$  are the \ac{RMS} delay spread and the delay proportionality factor. The per-cluster shadowing term is denoted by $Z_{c,k}\sim \textrm{Lognorm}(0,\sigma_Z^2)$.
These powers are then normalized  sich that $\sum_c \rho_c^\bs = \sum_c \rho_{c,k}^\ue=\sum_c \rho_{c,k}^\textrm{J}=1,\, \forall k$.
We propose the following coupling matrix 
\begin{customprop}{2}\label{prop omega}
    Let $\uu_{\bs,l}$ and $\uu_{\ue,k,m}$ denote the $l$-th largest eigenvector of $\CC_{\bs}$
 and the $m$-th largest eigenvector of $\CC_{\ue,k}$, then the geometry-based coupling matrix  
  \begin{align}
&\left[\OOmega_{k}\right]_{l,m}\hspace{-1mm}=\hspace{-1mm}\sum_{c=1}^{C}\,\rho_{c,k}^\textrm{J}\hspace{0.5mm} \nu_{c,l}\,\,\upsilon_{k,c,m}\\
&\nu_{c,l}=\left(\uu_{\bs,l}^\hermit\,\boldsymbol{C}_{\bs,c}\uu_{\bs,l}\right)/\Mbs \\
&\upsilon_{k,c,m}=\left(\uu_{\ue,k,m}^\hermit\,\boldsymbol{C}_{\ue,k,c}\uu_{\ue,k,m}\right)/\Mue
  \end{align}
\end{customprop}
This coupling matrix maps the propagation environment to eigenmode coupling: $\left[\OOmega_{k}\right]_{l,m}$ is large when both eigenmodes are excited by one or more clusters with high power. How much an eigenmode is excited by a cluster is measured by projecting the cluster contribution to the marginal covariance (e.g., $\mathcal{C}_{\bs,k,c}$) onto the corresponding eigenvector (e.g., $\uu_{\bs,k,l}$).
A Sinkhorn-like scaling is applied to $\OOmega$ \cite{sinkhorn1967concerning} to match the row and column sums to the BS and UE marginal eigenvalues, thus ensuring a fair comparison with $\OOmega^\textrm{kron}$.
\subsection{Temporal correlation}
In our previous works, the temporal correlation coefficient $\zeta$ followed a Jake's model. While analytically convenient, this model averages different propagation directions and does not account for the fact that the UE's Doppler shift is not isotropic, meaning that clusters at different positions age at different rates.
Let denote UE $k$ velocty with $\boldsymbol{\nu}_k = \norm{\boldsymbol{\nu}_k}[\cos(\theta_k),\sin(\theta_k)]$, with $\theta_k$ indicating the velocity's direction.
therefore, the Doppler shift seen by cluster $c$ is defined as $f_{\rmd,k,c}=(\overline{f}\norm{\boldsymbol{\nu}_k}/c_0)\cos(\varphi_{k,c}-\theta_k).$
Following the model presented in \cite[Lemma 1]{rivetti2026journal}, the correlation coefficient is 
\vspace{-2mm}
\begin{align}
\zeta_k(i)=\sum_{c=1}^C\rho_{k,c}^\ue\,\hspace{0.5mm}e^{j2\pi i f_{\rmd,k,c}T}e^{-2(\pi T i \varsigma_{\rmd,k,c})^2}\hspace{-1mm}.
\end{align}
where $\varsigma_{\rmd,k,c}$ represents the Doppler spread of said cluster, which can be related to the propagation geometry according to the following lemma.
\begin{customlemma}{1}\label{lemma temp}
Let us consider a cluster seen by UE $k$ at an angle $\varphi_{k,c}$ with an angle spread of $\varsigma_{\ue,k,c}$.
Assuming a Gaussian intra-cluster angular deviation, the cluster Doppler spread as seen by UE $k$, denoted by $\varsigma_{\rmd,k,c}$, is defined as 
\begin{align}
\varsigma_{\rmd,k,c}^2 \hspace{-1mm}= \hspace{-0.5mm}\overline{f}_{\rmd,k}^2\Big[ \hspace{-0.5mm}0.5
    \left(
    1+\hspace{-0.5mm}A_{k,c}^2
    \cos(2\epsilon_{k,c})
    \right)
    \hspace{-0.5mm}-\hspace{-0.5mm}
    A_{k,c}
    \cos^{2}(\epsilon_{k,c})\Big]
\end{align}
where   $A_{k,c}=e^{-\varsigma_{\ue,k,c}^{2}}$ ,$\overline{f}_{\rmd,k}=\overline{f}\norm{\boldsymbol{\nu_k}}/c_0 $, and $\epsilon_{k,c}=\theta_k-\varphi_{k,c}$
\end{customlemma}
\begin{IEEEproof}
Each ray scattered by cluster $c$ is seen by \ac{UE} $k$ at an angle $\varphi_{k,c,n}=\varphi_{k,c} + \delta_{\ue,k,c,n}$, where $\delta_{\ue,k,c,n}\sim\mathcal{N}(0,\varsigma_{\ue,k,c}^2)$ represents the deviation from the nominal angle. Then, the Doppler shift of ray $n$ is 
$f_{\rmd,k,c,n}=\overline{f}_{\rmd,k}\cos(\epsilon_{k,c}+\delta_{\ue,n})$.
Keeping in mind that $\mathbb{E}[\cos(\delta_{\ue,k,c,n})]=\exp\{-\varsigma_{\ue,k,c}^2/2\}$ and using the trigonometric identity $\cos(a+b)=\cos(a)\cos(b)-\sin(a)\sin(b)$ we compute $\mathbb{E}[f_{\rmd,k,c,n}]=\overline{f}_{\rmd,k}\exp\{-\varsigma_{\ue,k,c}^2/2\}\cos(\epsilon_{k,c})$.
Furthermore, by using the identity $\cos(a)^2=(1+\cos(2a))/2$, we are able to compute $\mathbb{E}[f_{\rmd,k,c,n}^2]=0.5\overline{f}_{\rmd,k}^2(1+ \exp\{-2\varsigma_{\ue,k,c}^2\}\cos(2\epsilon_{k,c})).$
The lemma follows by computing $\varsigma_{\rmd,k,c}^2=\mathbb{E}[f_{\rmd,k,c,n}^2]-\mathbb{E}[f_{\rmd,k,c,n}]^2$
\end{IEEEproof}

\section{Channel Estimation and Spectral Efficiency}
Fig. \ref{pilot table} shows the adopted pilot transmission pattern. Both \ac{UL} and \ac{DL} pilots are scheduled to provide to \acp{UE} and \ac{BS} with estimated \ac{CSI} for data detection and precoding. Each frame is composed of $\Delta_\ul$ \ac{UL} data slots and $\Delta_\dl$ downlink ones, bringing the total number of data slots to $\Delta=\Delta_\ul+\Delta_\dl$.
The precoding matrices adopted during pilot transmission, denoted by $\WW_{0,k,n}$ $\FF_{0,n}$, are assumed to be fixed and known by both parties\footnote{This avoids introducing non-Gaussian effective channel estimation \cite{kama2024downlink}, which is beyond the scope of this work.}.
Let us now focus on the \ac{UL} phase: each UE  transmits its pilot matrix $\beth_k \in \mathbb{C}^{\Mue \times \tau_\rmp }$ over $\tau_\rmp$ subcarriers: these matrices are mutually orthogonal, i.e., $\boldsymbol{\beth}_k\boldsymbol{\beth}_{k'}^\hermit=\tau_\rmp^\ul\mathbf{I} \leftrightarrow k=k'$.
The considerations that follow apply to all blocks; we thus omit the index $n$ for brevity.
Let us consider the arbitrary pilot slot $\iota$, the observed pilot matrix is  $  \YY_{\iota} = \sum_{k=1}^K \alpha_k\HH_{k,\iota}\WW_{0,k}\boldsymbol{\beth}_k
    + \NN_\iota\in \mathbb{C}^{\Mbs \times \tau_\rmp^\ul}$.
After pilot de-spreading, the vectorized observation is defined as 
\begin{align}
    \widetilde{\yy}_{\iota,k}=\hspace{-1mm}
\alpha_k\sqrt{\tau_\rmp^\ul} (\WW_{0,k}^\top \otimes \mathbf{I}_{\Mbs}) \hh_{k,\iota} +  \vecc(\widetilde{\NN_\iota})
\end{align}
The receiver uses $p$ pilot observations before the present one, bringing the total number of observations used for channel estimation to $p_\textrm{tot}=p+1$: We adopt the \ac{MMSE} channel estimation presented in \cite{rivetti2026cram}, that is
\begin{align}\label{h hat}
    \widehat{\hh}_{k,i} = \alpha_k \sqrt{\tau_\rmp^\ul}\hspace{1mm}\EE_{k,i}\WW_{\ptot,k}^\hermit\hspace{0.5mm}\overline{\AAb}_{k,i}^{-1}\hspace{0.5mm}\widetilde{\yy}^\ptot_{k}
\end{align}
where $\overline{\AAb}_{k,i}=\alpha_k^2\tau_\rmp^\ul\WW_{\ptot,k}^\hermit\MM_k\WW_{\ptot,k}^\hermit + \sigma^2_k\mathbf{I}_{\ptot\Mue\Mbs}$, $\WW_{\ptot,k} = \mathbf{I}_{\ptot} \otimes (\WW_{0,k}^\top \otimes \mathbf{I}_\Mbs)$, $\EE_{k,i}=\left[\zeta(i-\iota),\dots,\zeta(i-\iota+p(\Delta+1))  \right] \otimes \CC_k^\ul$, $\MM_{k}=\mathcal{T}\left([\zeta(0),\dots,\zeta(p(\Delta+1))]\right)\otimes \CC_k^\ul$, $\widetilde{\yy}^{\ptot}_k\hspace{-1mm}=
    [\widetilde{\yy}_{k,\iota }^\top,\dots\widetilde{\yy}_{k,\iota-p(\Delta+1)}^\top]^\top$.
The channel estimate $\widehat{\hh}_{k,i}$ is  $\widehat{\hh}_{k,i}\sim \mathcal{CN}(\mathbf{0},\widehat{\boldsymbol{\Xi}}_{k,i}=\alpha_k\tau_\rmp^\ul\EE_{k}\WW_{\ptot,k}^\hermit \overline{\AAb}_{k,i}^{-1}\WW_{\ptot,k}\EE_{k,i}^{\hermit}  )$. On the other hand, The channel estimation error $\widetilde{\hh}_{k,i}=\hh_{k,i}-\widehat{\hh}_{k,i}$, is distributed as $\widetilde{\hh}_{k,i}\sim \mathcal{CN}(\boldsymbol{0}, \widetilde{\boldsymbol{\Xi}}_{k,i}=\CC_k^\ul -\widehat{\boldsymbol{\Xi}}_{k,i})$.
We now derive achievable \ac{UL} and \ac{DL} \ac{SE}, assuming imperfect CSI and known precoders. 
We consider a generic subcarrier $v$ belonging to the block $n$. The BS adopts the arbitrary combining matrix $\TT_{k,i,n}\in \mathbb{C}^{\Mbs \times \Mue}$ for the detection of $\boldsymbol{x}_{k,i,v}$, while UE $k$ adopts the combining matrix 
$\QQ_{k,i,n}\in \mathbb{C}^{\Mue \times \Mue}$ to detect $\xx_{k,i,v}$.
assuming successive interference cancellation, \ac{UE}'s $k$ \ac{UL} \ac{SE} at data symbol time $i$ is \cite[Cor. 1]{debbah2022uplink}
  \begin{align}\label{SE}
   &{\rm{SE}}_{k,i,v}^\ul = \mathbb{E}\left\{ \log_2 \left\vert 
   \mathbf{I}_{\Mue} + \alpha_k^2\, \mathcal{H}_{k,i,n}^\hermit  \,
\boldsymbol{\Lambda}_{k,i,n}^{-1}\,\mathcal{H}_{k,i,n}
   \right\vert           
   \right\}.
\end{align} 
where $\mathcal{H}_{k,i,n}=\TT_{k,i,n}^\hermit\widehat{\HH}_{k,i,n}\WW_{k,i,n}$ and $\boldsymbol{\Lambda}_{k,i,n}$ is defined in \eqref{multipage 2}.
\begin{figure*}[!t]
    \begin{align}\label{multipage 2}
        \boldsymbol{\Lambda}_{k,i,n}= \TT_{k,i,n}^\hermit\Big(\sum\nolimits_{l=1,\, l \neq k}^K\hspace{-2mm} \alpha_l^2\widehat{\HH}_{l,i,n} \WW_{l,i,n}\WW_{l,i,n}^\hermit\widehat{\HH}_{l,i,n}^\hermit + 
\sum\nolimits_{l=1}^K \alpha_l^2 \textrm{Cov}\big[\widetilde{\HH}_{l,i,n} \WW_{l,i,n}\big]  +\sigma^2\mathbf{I}_{\Mbs}\Big)\TT_{k,i,n}
    \end{align}
    \vspace{-5mm}
        \begin{align}\label{multipage 3}
        \boldsymbol{\Gamma}_{k,i,n}= \QQ_{k,i,n}^\hermit\Big(\sum\nolimits_{l=1,\, l \neq k}^K \alpha_k^2\widehat{\HH}_{k,i,n,l}^\dl \FF_{l,i,n}\FF_{l,i,n}^{\hermit}\left(\widehat{\HH}_{k,i,n}^{\dl}\right)^\hermit  + 
 \alpha_k^2 \textrm{Cov}\big[\widetilde{\HH}_{k,i,n}^\dl \FF_{i,n}\big]  +\sigma^2\mathbf{I}_{\Mue}\Big)\QQ_{k,i,n}
    \end{align}
\noindent\raisebox{2mm}{\rule{\textwidth}{0.4pt}}
\end{figure*}
Here $\widetilde{\HH}_{k,i,n}$ has been obtained by stacking $\widetilde{\hh}_{k,i,n}$ into a matrix.
The \ac{DL} channel estimator is not explicitly reported for brevity, as it follows the same covariance-based approach of \eqref{h hat} and is reported in \cite{rivetti2026journal}. Assuming known BS precoders at the UEs, the \ac{DL} SE can be modelled as 
\begin{align}
      &{\rm{SE}}_{k,i,v}^\dl\hspace{-1mm}=\hspace{-1mm} \mathbb{E}\left\{ \log_2 \left\vert 
   \mathbf{I}_{\Mue}\hspace{-1mm} + \alpha_k^2\hspace{-1mm}\left(\mathcal{H}_{k,i,n}^{\dl}\right)^\hermit 
\boldsymbol{\Gamma}_{k,i,n}^{-1}\mathcal{H}_{k,i,n}^\dl
   \right\vert           
   \right\}.
\end{align}
where $\mathcal{H}_{k,i,n}^\dl=\QQ_{k,i,n}^\hermit\widehat{\HH}_{k,i,n}^\dl\FF_{k,i,n}$ and $\boldsymbol{\Gamma}_{k,i,n}$ is defined in \eqref{multipage 3}.
Therefore, the average \ac{UL} \ac{SE}, and analogously the DL SE, over the frame, aggregated over the subcarriers and UEs, is defined as $\textrm{SE}_\ul=\sum_{k=1}^K
\sum_{i=1}^{\Delta_\ul}\sum_{v=1}^V \textrm{SE}_{k,i,v}^\ul\,/(\Delta+2)$. 

\subsection{Precoding strategy}
\vspace{-1mm}
The adopted precoding strategy differs between data slots, where CSI is available, and pilot slots, where the previous estimate has drifted too far from its actual value. In the absence of CSI, the BS implements eigen-beamforming, that is $\FF_{0,n,k}=[\uu_{\bs,k,1},\dots,\uu_{\bs,k,\Mue}]$, while $\WW_{0,k,n}=\mathbf{I}_{\Mue}$.
In the presence of CSI, the BS adopts MMSE precoding \cite[Ch.~5]{demir2021foundations} while the UEs precoders are $\WW_{k,i,n}=\boldsymbol{U}_{k,i,n}$, where $\boldsymbol{U}_{k,i,n}$ is the left singular matrix of $\widehat{\HH}_{k,i,n}^\dl$.
Precoders are normalized to ensure an equal power allocation between subcarriers and streams.
We compare \ac{ZF} and \ac{MMSE} combining at the BS: the ZF combining matrix is $\TT_{k,i,n}^\textrm{ZF}=\HH_{i,n}^\dagger$, with $\HH_{i,n}=[\alpha_1\HH_{1,i,n}, \alpha_K\HH_{K,i,n}]$,while the MMSE combining matrix  $\TT_{k,i,n}^\textrm{MM}$ is defined in \eqref{Tmm} .  The UEs adopt the combining matrix $\QQ_{k,i,n}$, defined in \eqref{multipage 3}.
\begin{figure*}[!t]
\vspace{-7mm}
    \begin{align}\label{Tmm}
\TT_{k,i,n}^\textrm{MM}\hspace{-1mm}=\hspace{-1mm}\Big(\sum\nolimits_{k=1}^K \alpha_k^2 \Big(\widehat{\HH}_{k,i,n}\WW_{k,i,n}\WW_{k,i,n}^\hermit\widehat{\HH}_{k,i,n}^\hermit  + \textrm{Cov}[\widetilde{\HH}_{k,i,n}^\top\WW_{k,i,n}] \Big)+\sigma^2\mathbf{I}_{\Mbs} \Big)^{-1} \hspace{-1mm}\alpha_k\widehat{\HH}_{k,i,n}\WW_{k,i,n}
\end{align}. 
\vspace{-5mm}
\begin{align}\label{Qmm}
\QQ_{k,i,n}=\hspace{-1mm}\Big(\alpha_k^2 \widehat{\HH}_{k,i,n}^\dl\FF_{i,n}\FF_{i,n}^\hermit \left(\widehat{\HH}_{k,i,n}^\dl\right)^\hermit  + \textrm{Cov}[\widetilde{\HH}_{k,i,n}^\dl\FF_{i,n}] +\sigma^2\mathbf{I}_{\Mbs} \Big)^{-1} \hspace{-1mm}\alpha_k\widehat{\HH}_{k,i,n}^\dl\FF_{i,n}
\end{align}
\noindent\raisebox{2mm}{\rule{\textwidth}{0.4pt}}
\vspace{-9mm}
\end{figure*}
\vspace{-1mm}
{\scriptsize
\begin{table}[t]\label{table sim}
\begin{center}
\caption {\centering Simulation parameters default values}
\begin{tabular}{|m{0.5\linewidth} | m{0.4\linewidth}|} 
 \hline
 \textbf{Parameter} & \textbf{value}\\
 \hline\hline 
 BS and UE antennas ($\Mbs,\, \Mue$)  & $16,\, 4$\\
 \hline
UE positions ($\ee_1, \ee_2$) & $300[\cos(\pm 40^\circ), \sin(\pm 40^\circ)]\,$m\\
 \hline
 UE pathloss ($\alpha_k$)& 3GPP street canyon model \cite{european2018study}\\
 \hline
UE velocities ($\boldsymbol{\nu}_1, \boldsymbol{\nu}_2$) & $10[\cos(0), \sin(0)]$ m/s\\
 \hline
Frequency blocks,  ($B$)  & $2$\\
  \hline
Subc. per block ($V_\textrm{cho}$)  & $20$\\
  \hline
  \vspace{0.5mm}
Carrier frequency, Symbol time ($\overline{f},T$)  & $4$\,GHz, $53\,\mu$s\\
 \hline
UL and DL data slots ($\Delta_\ul,\Delta_\dl$)  & $5$\\
\hline
UL and DL pilot length ($\tau^\ul_\rmp,\tau^\dl_\rmp$)  & $20$\\
 \hline
BS and UEs total power ($P_\bs, P_\ue$) & $30,\,20$ dBm\\
 \hline
 Number of clusters ($C$) & $25$\\
  \hline
 $c$-th cluster effective radius ($r_c$) & $r_c\sim\textrm{Lognorm}(\log(8),0.5)$\\
   \hline
 RMS delay spread ($DS$) & $150\,\mu$s\\
  \hline
 proportionality constant ($r_\tau$) & $2.4$\\
   \hline
 Shadowing variance ($\sigma_Z^2$) & $3$\\
    \hline
 UEs and BS noise variance ($\sigma_\bs^2,\, \sigma_k^2$) & $-150\,$dB\\
 \hline
 \end{tabular}
 \end{center}
\end{table}
}

\begin{figure}[t!]
\centering
\begin{minipage}{0.8\textwidth}
\input{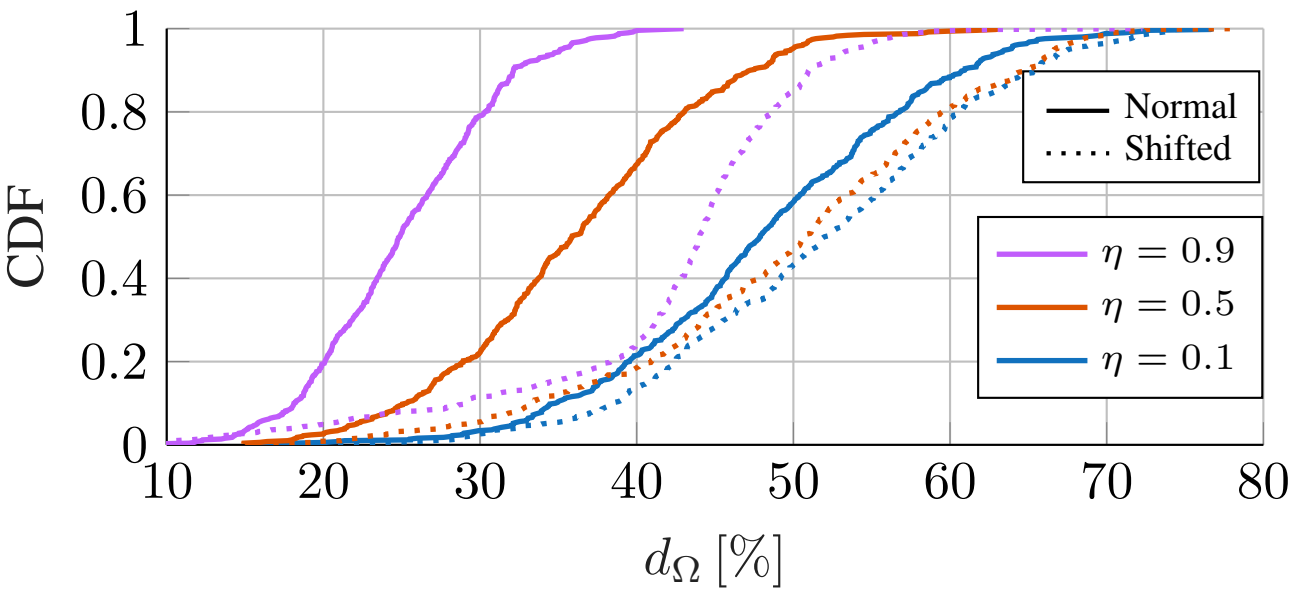}
\end{minipage}
\caption{CDF of the covariance mismatch between the Kronecker separable model and the Jointly correlated one. The CDF is computed over 500 independent geometric setups.} 
\label{fig cdf}
\vspace{-4mm}
\end{figure}


\section{Numerical Results}
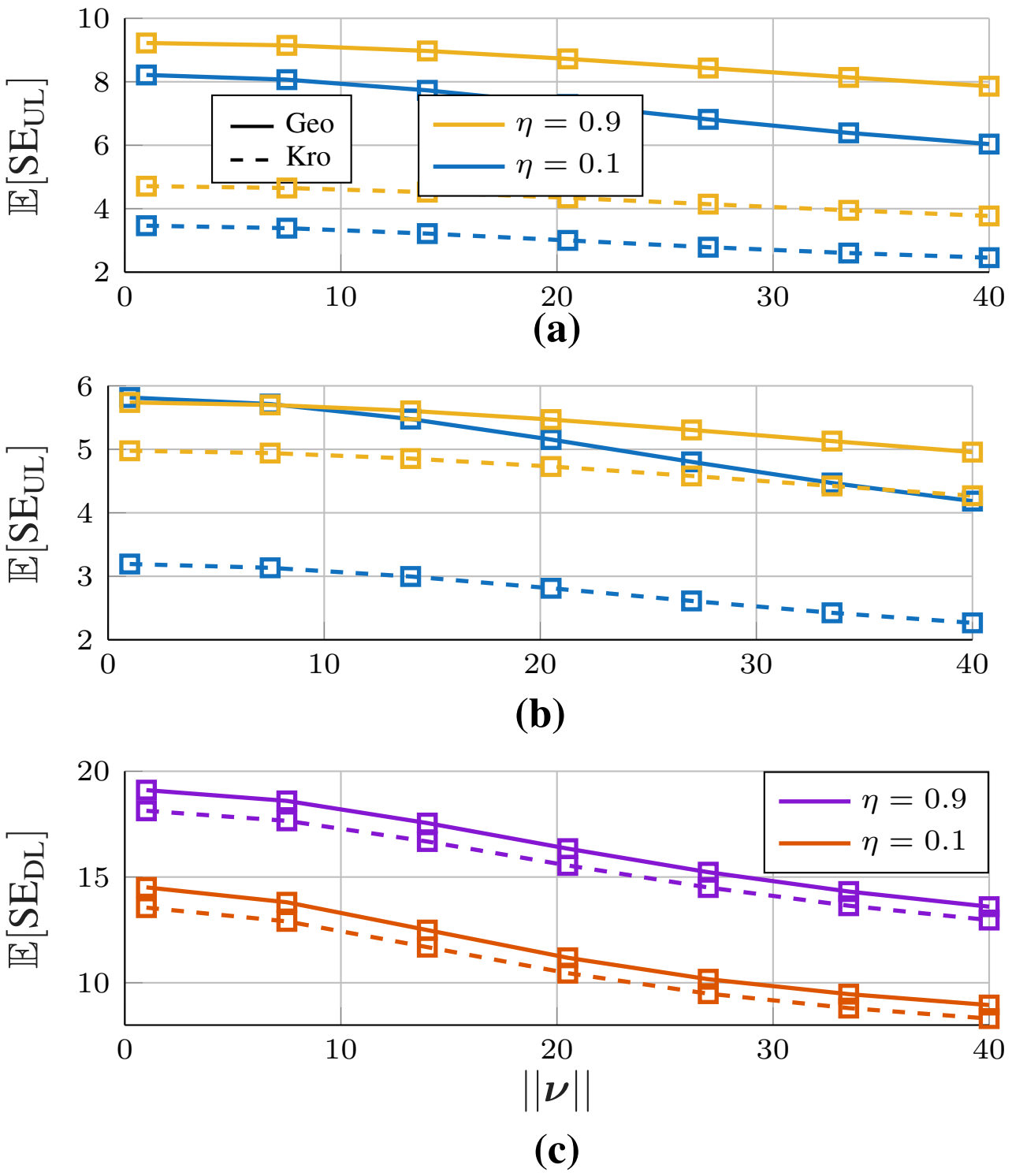
\begin{figure}[t!]
\centering
\begin{minipage}{0.8\textwidth}
%
%
\definecolor{mycolor1}{rgb}{0.06600,0.44300,0.74500}%
\definecolor{mycolor2}{rgb}{0.92900,0.69400,0.12500}%
\definecolor{mycolor3}{rgb}{0.52100,0.08600,0.81900}%
\definecolor{mycolor4}{rgb}{0.12941,0.12941,0.12941}%
\begin{tikzpicture}

\begin{axis}[%
width=0.5\textwidth,
height=0.83859in,
at={(0.97in,0.563in)},
scale only axis,
clip=false,
xmin=0,
xmax=40,
xlabel style={font=\color{mycolor4}},
ticklabel style={font=\scriptsize},
ymin=2,
ymax=10,
ylabel style={font=\color{mycolor4}},
ylabel={$\mathbb{E}[\textrm{SE}_\ul]$},
axis background/.style={fill=white},
axis x line*=bottom,
axis y line*=left,
xmajorgrids,
ymajorgrids,
legend style={at={(0.6,0.7)},legend cell align=left, align=left,font=\scriptsize}
]

\addplot [color=mycolor1, line width=1.0pt, mark=square, mark options={solid, mycolor1},forget plot]
  table[row sep=crcr]{%
1	8.21467870103827\\
7.5	8.06685819569939\\
14	7.73032961189653\\
20.5	7.28459470160226\\
27	6.81596048845635\\
33.5	6.38930540149145\\
40	6.03659724780465\\
};

\addplot [color=mycolor1, dashed, line width=1.0pt, mark=square, mark options={solid, mycolor1}, forget plot]
  table[row sep=crcr]{%
1	3.46392252780518\\
7.5	3.3863708472681\\
14	3.2156727066512\\
20.5	3.00073322023424\\
27	2.78689077406841\\
33.5	2.60212650591783\\
40	2.45609311507139\\
};

\addplot [color=mycolor2, line width=1.0pt, mark=square, mark options={solid, mycolor2},forget plot]
  table[row sep=crcr]{%
1	9.22229690218836\\
7.5	9.14792051129436\\
14	8.97184367871872\\
20.5	8.72185046834167\\
27	8.43285079501384\\
33.5	8.13761357469645\\
40	7.86043805687138\\
};

\addplot [color=mycolor2, dashed, line width=1.0pt, mark=square, mark options={solid, mycolor2}, forget plot]
  table[row sep=crcr]{%
1	4.7089746617224\\
7.5	4.6522856926217\\
14	4.52123765923391\\
20.5	4.34194271930074\\
27	4.14316780329142\\
33.5	3.94814763436636\\
40	3.77147692709045\\
};

 \node[
    draw=black,
    fill=white,
    font=\scriptsize,
    anchor=north west,
    align=left,
    inner sep=4pt
] at (rel axis cs:0.1,0.7) {
\raisebox{0.3ex}{\tikz \draw[thick,solid] plot coordinates {(0,0) (2,0)};}~Geo\\
\raisebox{0.3ex}{\tikz \draw[thick,dashed] plot coordinates {(0,0) (2,0)};}~Kro
};

\addlegendimage{ mycolor2, line width=1pt}
\addlegendentry{$\eta=0.9$}
\addlegendimage{ mycolor1, line width=1pt}
\addlegendentry{$\eta=0.1$}

\node at (axis description cs:0.5,-0.23) {\textbf{(a)}};

\end{axis}
\end{tikzpicture}%
\end{minipage}
\begin{minipage}{0.8\textwidth}
%
%
\definecolor{mycolor1}{rgb}{0.06600,0.44300,0.74500}%
\definecolor{mycolor2}{rgb}{0.92900,0.69400,0.12500}%
\definecolor{mycolor3}{rgb}{0.52100,0.08600,0.81900}%
\definecolor{mycolor4}{rgb}{0.12941,0.12941,0.12941}%
\begin{tikzpicture}

\begin{axis}[%
width=0.5\textwidth,
height=0.839in,
at={(0.97in,0.563in)},
scale only axis,
clip=false,
xmin=0,
xmax=40,
xlabel style={font=\color{mycolor4}},
ticklabel style={font=\scriptsize},
ymin=2,
ymax=6,
ylabel style={font=\color{mycolor4}},
ylabel={$\mathbb{E}[\textrm{SE}_\ul]$},
axis background/.style={fill=white},
axis x line*=bottom,
axis y line*=left,
xmajorgrids,
ymajorgrids,
legend style={at={(0.6,0.6)},legend cell align=left, align=left,font=\scriptsize}
]

\addplot [color=mycolor1, line width=1pt, mark=square, mark options={solid, mycolor1}]
  table[row sep=crcr]{%
1	5.81560998553393\\
7.5	5.71336175056672\\
14	5.47673092504223\\
20.5	5.15421904879732\\
27	4.8025779905389\\
33.5	4.46973617114878\\
40	4.18461039073907\\
};

\addplot [color=mycolor1, dashed, line width=1.0pt, mark=square, mark options={solid, mycolor1}, forget plot]
  table[row sep=crcr]{%
1	3.19312060896845\\
7.5	3.13363969284131\\
14	2.99669845309825\\
20.5	2.81161797334699\\
27	2.61184691340361\\
33.5	2.42477386143929\\
40	2.26626255263085\\
};



\addplot [color=mycolor2, line width=1.0pt, mark=square, mark options={solid, mycolor2}]
  table[row sep=crcr]{%
1	5.73837592208056\\
7.5	5.69935982638586\\
14	5.60570034483076\\
20.5	5.46914988971031\\
27	5.30515277199688\\
33.5	5.12957115841472\\
40	4.95600447740575\\
};

\addplot [color=mycolor2, dashed, line width=1.0pt, mark=square, mark options={solid, mycolor2}, forget plot]
  table[row sep=crcr]{%
1	4.97704006162072\\
7.5	4.94086897083458\\
14	4.85442124866395\\
20.5	4.72933779620711\\
27	4.58057627928765\\
33.5	4.42303591517521\\
40	4.26901290751563\\
};


\node at (axis description cs:0.5,-0.3) {\textbf{(b)}};

\end{axis}
\end{tikzpicture}%
\end{minipage}
\begin{minipage}{0.8\textwidth}
%
%
\definecolor{mycolor1}{rgb}{0.86600,0.32900,0.00000}%
\definecolor{mycolor2}{rgb}{0.52100,0.08600,0.81900}%
\definecolor{mycolor3}{rgb}{0.12941,0.12941,0.12941}%
\begin{tikzpicture}

\begin{axis}[%
width=0.5\textwidth,
height=0.83859in,
clip=false,
at={(0.97in,0.563in)},
scale only axis,
xmin=0,
xmax=40,
xlabel style={font=\color{mycolor3}},
xlabel={$||\boldsymbol{\nu}||$},
ticklabel style={font=\scriptsize},
ymin=8,
ymax=20,
ylabel style={font=\color{mycolor3}},
ylabel={$\mathbb{E}[\textrm{SE}_\dl]$},
axis background/.style={fill=white},
axis x line*=bottom,
axis y line*=left,
xmajorgrids,
ymajorgrids,
x label style={
        at={(axis description cs:0.5,-0.4)},
        anchor=south
    },
legend style={at={(1,1)},legend cell align=left, align=left,font=\scriptsize}
]

\addplot [color=mycolor1, line width=1.0pt, mark=square, mark options={solid, mycolor1}, forget plot]
  table[row sep=crcr]{%
1	14.5105547573123\\
7.5	13.8067952936861\\
14	12.4913762539235\\
20.5	11.181680713235\\
27	10.1680166005228\\
33.5	9.4626511490244\\
40	8.94731563012666\\
};

\addplot [color=mycolor1, dashed, line width=1.0pt, mark=square, mark options={solid, mycolor1}, forget plot]
  table[row sep=crcr]{%
1	13.5520250118971\\
7.5	12.9031571162648\\
14	11.685032127579\\
20.5	10.4550923692744\\
27	9.48574021050469\\
33.5	8.80298426795992\\
40	8.30782996976344\\
};

\addplot [color=mycolor2, line width=1.0pt, mark=square, mark options={solid, mycolor2}, forget plot]
  table[row sep=crcr]{%
1	19.1075218203412\\
7.5	18.5996253511013\\
14	17.551471143337\\
20.5	16.3413564692046\\
27	15.2281905832826\\
33.5	14.3137400156442\\
40	13.5973512174872\\
};
\addplot [color=mycolor2, dashed, line width=1.0pt, mark=square, mark options={solid, mycolor2}, forget plot]
  table[row sep=crcr]{%
1	18.1307919995137\\
7.5	17.6569108580505\\
14	16.6787624571041\\
20.5	15.5478459653307\\
27	14.5042626307472\\
33.5	13.643227660658\\
40	12.9656854731191\\
};
\addlegendimage{ mycolor2, line width=1pt}
\addlegendentry{$\eta=0.9$}
\addlegendimage{ mycolor1, line width=1pt}
\addlegendentry{$\eta=0.1$}

\node at (axis description cs:0.5,-0.5) {\textbf{(c)}};

\end{axis}

\end{tikzpicture}%
\end{minipage}
\vspace{-2mm}
\caption{Average SE in the UL, for both MMSE (a) and ZF (b) BS combining, and in the DL (c). We obtain the reported curves by averaging 20 Monte Carlo realizations of the propagation environment; for each environment, we average the \ac{SE} over 500 channel and precoder realizations.} 
\label{fig SE}
\end{figure}

\vspace{-1mm}
The default value of the simulation parameters is reported in Table I. The effective radii are drawn independently from the log-normal distribution shown in Table I : $r_c$ is clipped between $2$ and $30\,$m to avoid unrealistically large radii. The resulting angle spreads are then also clipped between $1^\circ$ and $10^\circ$.
The number of clusters is divided into $C_\textrm{dom}=\lceil\eta C \rceil$ dominant clusters and $C_\textrm{back}=C-C_\textrm{dom}$ background one : this allows us to control the propagation environment by tuning $\eta$. The dominant clusters are divided among the UEs as $C_\textrm{dom}^k=\lceil C_\textrm{dom}/K \rceil$ and their position is randomly generated as $\pp_{c}^{\textrm{dom},k}=\beta_{k,c}\ee_k + \kappa_{k,c}\vv_k^{\perp}$. Here $\beta_{k,c}\sim \mathcal{U}(0.35,0.65)$, $\kappa_{k,c}\sim\mathcal{N}(0,20)\,$m while $\vv_k^{\perp}$ is a unit vector orthogonal to $\ee_k$. On the other hand, the background clusters positions $\pp_c^\textrm{back}=[p_{c,x}^\textrm{back},p_{c,y}^\textrm{back}]^\top$, with $p_{c,x}^\textrm{back}\sim \mathcal{U}(0,225)\,$, and $p_{c,y}^\textrm{back}\sim \mathcal{U}(-300,300)\,$.
The proposed Weischelberger model is denoted by "Geo'', while the Kronecker-separable baseline is labeled as 
"Kron''. 
Fig~\ref{fig cdf} shows the empirical \ac{CDF} of the covariance mismatch between the Kronecker separable model and the jointly correlated one: the solid curves show the nominal geometry while the dashed ones represent setups where the dominant-clusters centroid is shifted by $30^\circ$. We observe that the accuracy of a Kronecker-separable approximation is strongly favored by a concentrated scattering environment (i.e., $\eta=0.9$), exhibiting a $50\%$ probability of being below $25\%$. A dominant-cluster centroid shift makes the Kronecker separable approximation much less accurate and less dependent on $\eta$, as the gap between the dotted curves shrinks substantially. This indicates that the Kronecker accuracy is favored by a strongly concentrated propagation environment where the clusters lie on the  line between \ac{BS} and \ac{UE}.
We now move to investigating the impact of a Kronecker-separable approximation on the achievable \ac{SE}: Fig.~\ref{fig SE} shows the average UL and DL \acp{SE} evolution for increasing UE mobility in diffusive (i.e. $\eta=0.1$) and concentrated (i.e. $\eta=0.9$) scattering conditions.  
Figs \ref{fig SE}a and \ref{fig SE}b show the UL SE with MMSE and ZF BS combining, respectively, while Fig. \ref{fig SE}c shows the DL SE.
Increasing
$\norm{\boldsymbol{\nu}}$ degrades all the SEs due to an increased aging rate of the channel.A  concentrated scattering environment is generates higher SEs: concentrated scattering geometries produce smaller Doppler spreads and their  $\zeta_k(\cdot)$ is less sensitive to UE mobility.
The DL SE in Fig.~\ref{fig SE}c is
more sensitive to $\norm{\boldsymbol{\nu}}$ than the UL, as the UL SE benefits from centralized BS-side detection, and shows a low-to-moderate Geo-Kron gap.
On the contrary, the UL SE is less sensitive to UE mobility and exhibits a large Geo-Kron gap. Interestingly, with ZF BS combining, the Geo-Kron gap becomes smaller in concentrated propagation environments.

\section{Conclusions}
We proposed a geometry-aware channel model where the same clustered propagation environment determines both the Weischelberger eigen-mode coupling and the temporal correlation decay. Channel aging is now determined by the UE motion relative to the scattering geometry, rather than its velocity magnitude alone. Using explicit UL and DL pilots, we evaluated channel estimation and imperfect-CSI SE. The results show that Kronecker separability is an adequate approximation in highly concentrated propagation regimes, but can miss relevant eigenmode coupling in diffuse environments. Moreover, diffuse propagation induces stronger Doppler diversity, causing stronger channel aging and larger SE degradation at high UE velocities. This motivates eigenmode-coupled models for mobility-sensitive MIMO and ISAC systems.


\vspace*{-2mm}

\bibliographystyle{IEEEtran}
\bibliography{IEEEabrv,auxiliary/biblio,JCASv10}
\end{document}